\documentclass[article,prl,balancelastpage,onecolumn,showpacs]{revtex4}
\usepackage{makeidx}
\usepackage{amssymb}
\usepackage{dcolumn}
\usepackage{graphicx}
\usepackage{acronym}

\input{tcilatex}

\begin{document}

\title{Singular states of relativistic fermions in the field of a circularly
polarized electromagnetic wave and constant magnetic field}
\author{B. V. Gisin }
\affiliation{IPO, Ha-Tannaim St. 9, Tel-Aviv 69209, Israel. E-mail:
borisg2011@bezeqint.net}
\date{}

\begin{abstract}
Dirac's equation in the field of a circularly polarized electromagnetic wave
and constant magnetic field has exact localized non-stationary solutions.
The solutions corresponds relativistic fermions only. Among them singular
solutions with energy eigenvalues close to each other are found. The
solutions are most practicable and can be separated by means of the phase
matching between the momentum of the electromagnetic wave and spinor.
Characteristic parameters of the singular states are defined.\hfill
\end{abstract}

\pacs{03.65.Ge, 71.70.Di, 13.49.Em; \hspace{20cm}}
\maketitle

\section{1. Introduction}

Recently a new class of exact localized non-stationary solutions to Dirac's
equation in the field of a traveling circularly polarized electromagnetic
wave and constant magnetic field was presented \cite{S1}. These solutions
correspond to stationary states (Landau levels) in a rotating and co-moving
frame of references. In contrast to classical case, the wave function of
such solutions never can be presented as large and small two-component
spinor. They are relevant only to relativistic fermions.

In \cite{S1}, as an example, "ground state" of these solutions was
considered (quotation marks are used here because states are
non-stationary). The problem of the magnetic resonance was studied in a
general manner in the framework of the classical field theory \cite{LTP} and
an exact expression for oscillations of the magnetic moment found. It was
shown that in the condition of the magnetic resonance, i.e., at the
amplitude maximum in respect to the constant magnetic field, this amplitude
still remains dependent on the momentum, frequency and amplitude of the
electromagnetic wave. Some evaluations of parameters for practical
realization of such solutions were performed.

In the given paper the consideration is expanded to "excited states".
Moreover, singular solutions are found. These solutions have some
interesting features applicable to practice. In particular, the phase
matching between the momentum of the electromagnetic wave and spinor may
enhance the precision technic \cite{mag} for measurements of the fermion
mass and magnetic moment.

For simplicity, the consideration is restricted by "pure quantum mechanics"
without reference to quantum field theory \cite{quant}, \cite{bog}.

\section{2. States of Dirac's equation in rotating electromagnetic field}

We consider Dirac's equation\ 
\begin{equation}
i\hbar \frac{\partial }{\partial t}\Psi =c\mathbf{\alpha }(\mathbf{\mathbf{p}%
}-\frac{e}{c}\mathbf{\mathbf{A}})\Psi +\beta mc^{2}\Psi =0  \label{Dir}
\end{equation}%
in the electromagnetic field with potential $A_{x}=-\frac{1}{2}H_{z}y+\frac{1%
}{k}H\cos (\Omega t-kz),$ $A_{y}=\frac{1}{2}H_{z}x+\frac{1}{k}H\sin (\Omega
t-kz),$ where $k=\varepsilon \Omega /c$ is the propagation constant, $\Omega 
$ is the frequency, the sign change of $\Omega $ corresponds to the opposite
polarization, values $\varepsilon =1$ and $\varepsilon =-1$ are used when
the wave propagates along the $z$-axis and opposite direction respectively, $%
c$ is the speed of light, $\alpha _{k},\beta $ are Dirac's matrices, $H$ is
the amplitude of this wave. This potential corresponds a plane circularly
polarized wave and constant magnetic field.

Stationary solutions of Eq. (\ref{Dir}) exist in a rotating and co-moving
frame of reference. Coordinates of the frame are%
\begin{eqnarray}
\tilde{x} &=&r\cos \tilde{\varphi},\text{ \ }\tilde{y}=r\sin \tilde{\varphi}%
,\   \label{xy} \\
\tilde{\varphi} &=&\varphi -\Omega t+kz,\text{ }\tilde{t}=t,\text{ }\tilde{z}%
=z,  \label{phi}
\end{eqnarray}%
where the tilde is used for these coordinates. The transformation $\ \tilde{%
\Psi}=\exp [\frac{1}{2}\alpha _{1}\alpha _{2}(\Omega t-kz)]\Psi $ \
describes the translation of a spinor into this rotating reference frame.
Obviously the operator 
\begin{equation}
\hbar (\frac{\partial }{\partial t}+\text{\ }\varepsilon c\frac{\partial }{%
\partial z})  \label{inxy}
\end{equation}%
is invariant under the transformation (\ref{xy}).

We use constants $E$ and $p$ as "energy" and "momentum along the $z$-axis"
for stationary states in the rotating frame.\ Once these states are found,
the wave function as well as coordinates are translated back into the
initial (non-rotating) frame. In this frame the operator (\ref{inxy}), in
contrast to operators of energy and momentum, commutes with the Hamiltonian
of Eq. (\ref{Dir}).

The Dirac equation has exact localized non-stationary solutions $\Psi =\exp
[-iEt/\hbar +ipz/\hbar -\alpha _{1}\alpha _{2}(\Omega t-kz)/2+D]\psi ,$ $%
D=-d(\tilde{x}^{2}+\tilde{y}^{2})/2+d_{1}\tilde{x}+d_{2}\tilde{y},$ were $%
\psi $ is a spinor.\ States may be classified in accordance with the form of
this spinor $\psi $. A constant spinor describes "ground state". A spinor
polynomial in $\tilde{x}$,\ $\tilde{y}$ corresponds to "excited states".

Localized solutions exist if the parameter $d$ is positive and defined by
the equality%
\begin{equation}
d=|\frac{eH_{z}}{2\hbar c}|.  \label{d2}
\end{equation}%
In accordance with Eq. (\ref{d2}), two types of solutions are possible. We
denote them as $\psi _{-}$ \ for $eH_{z}<0$ and $\psi _{+}$ for $eH_{z}>0.$
These normalized spinors of the ground state have the form

\begin{equation}
\psi _{-0}=N_{-}\left( 
\begin{array}{c}
h\mathcal{E} \\ 
-\varepsilon (\mathcal{E}+1)(\mathcal{E}-\mathcal{E}_{0}) \\ 
\varepsilon h\mathcal{E} \\ 
-(\mathcal{E}-1)(\mathcal{E}-\mathcal{E}_{0})%
\end{array}%
\right) ,\text{ \ }\psi _{+0}=N_{+}\left( 
\begin{array}{c}
(\mathcal{E}+1)(\mathcal{E}+\mathcal{E}_{0}) \\ 
\varepsilon \mathcal{E}h \\ 
-\varepsilon (\mathcal{E}-1)(\mathcal{E}+\mathcal{E}_{0}) \\ 
-\mathcal{E}h%
\end{array}%
\right) ,  \label{psinp}
\end{equation}%
where $N_{\mp }$ is defined by the normalization condition $\int \Psi ^{\ast
}\Psi dxdy=1.$

\begin{eqnarray}
N_{\mp } &=&\frac{\sqrt{d/2\pi }\exp (-d_{2}^{2}/2d)}{\sqrt{(\mathcal{E}%
^{2}+1)(\mathcal{E}\mp \mathcal{E}_{0})^{2}+h^{2}\mathcal{E}^{2}}},
\label{N0pm} \\
d_{1} &=&\mp id_{2},\text{ \ }d_{2}=\frac{\mathcal{E}_{0}mch}{2\hbar (%
\mathcal{E}\mp \mathcal{E}_{0})}.  \label{dd12}
\end{eqnarray}%
The upper and lower sign before a parameter corresponds to solutions with
negative and positive $eH_{z}$ respectively.

The normalized eigenvalue of the operator (\ref{inxy}) is $\mathcal{E}\equiv
(E-\varepsilon pc)/mc^{2}$. Eigenvalues of the "ground state" obey the
characteristic equation%
\begin{equation}
-\mathcal{E}(\mathcal{E}+\Lambda _{\mp })+1+\frac{\mathcal{E}}{\mathcal{E}%
\mp \mathcal{E}_{0}}h^{2}=0,  \label{E0}
\end{equation}%
where%
\begin{equation}
\mathcal{E}_{0}\mathcal{=}\frac{2\hbar d}{\Omega m},\text{ \ }\Lambda _{\mp
}=\frac{2\varepsilon pc\mp \hbar \Omega }{mc^{2}},\text{ \ \ }h=\frac{e}{%
kmc^{2}}H.  \label{vh}
\end{equation}

Obviously, wave functions (\ref{psinp}) cannot be presented as a small and
large two-component spinor. It means that the difference $E^{2}-m^{2}c^{2}$
cannot be small and these solutions correspond only to the relativistic case.

For the first "excite state" $\psi =\psi _{0}+\tilde{x}\psi _{x}+\tilde{y}%
\psi _{y}$, where $\psi _{0},\psi _{x},\psi _{y}$ are constant spinors.
There exist non-degenerative solutions with $\psi =\psi _{\mp 0}(1-id\tilde{x%
}/d_{2}\mp d\tilde{y}/d_{2}),$ where $\psi _{\mp 0}$ is defined by Eq. (\ref%
{psinp}) and the normalization coefficient%
\begin{equation}
N_{\mp 1}=\exp [-\frac{d_{2}^{2}}{2d}]\frac{d_{2}\sqrt{d}}{\sqrt{2\pi
(d+d_{2}^{2})}}\frac{1}{\sqrt{h^{2}\mathcal{E}^{2}+(\mathcal{E}^{2}+1)(%
\mathcal{E}\mp \mathcal{E}_{0})^{2}}}  \label{N1pm}
\end{equation}%
is used instead of $N_{\mp }$ . The characteristic equation coincides with
Eq. (\ref{E0}) if 
\begin{equation}
\Lambda _{\mp }=\frac{2\varepsilon pc\mp 3\hbar \Omega }{mc^{2}}\text{.}
\label{la}
\end{equation}%
Two types of degenerate solutions are 
\begin{eqnarray}
\psi _{-11} &=&N_{-11}\left( 
\begin{array}{c}
\varepsilon (\mathcal{E}+1) \\ 
h-i\frac{2\hbar d}{mc}\tilde{x}+\frac{2\hbar d}{mc}\tilde{y} \\ 
-(\mathcal{E}-1) \\ 
-\varepsilon \lbrack h-i\frac{2\hbar d}{mc}\tilde{x}+\frac{2\hbar d}{mc}%
\tilde{y}]%
\end{array}%
\right) ,  \label{s11} \\
N_{-11} &=&\exp [-\frac{d_{2}^{2}}{2d}]\frac{\sqrt{d}(\mathcal{E}-\mathcal{E}%
_{0})/\sqrt{2\pi }}{\sqrt{h^{2}\mathcal{E}^{2}+(\mathcal{E}-\mathcal{E}%
_{0})^{2}(\mathcal{E}^{2}+1+4\hbar ^{2}d/m^{2}c^{2})}},  \label{Nd11}
\end{eqnarray}%
\begin{eqnarray}
\psi _{-12} &=&N_{-12}\left( 
\begin{array}{c}
\lbrack \mathcal{E}(\mathcal{E}+\Lambda _{-})-1]+i\frac{\Omega }{c}h\mathcal{%
E}\tilde{x}-\frac{\Omega }{c}h\mathcal{E}\tilde{y} \\ 
-\varepsilon (\mathcal{E}+1)[h+i\frac{\Omega }{c}(\mathcal{E-E}_{0})\tilde{x}%
-\frac{\Omega }{c}(\mathcal{E-E}_{0})\tilde{y}] \\ 
\varepsilon \{[\mathcal{E}(\mathcal{E}+\Lambda _{-})-1]+i\frac{\Omega }{c}h%
\mathcal{E}\tilde{x}-\frac{\Omega }{c}h\mathcal{E}\tilde{y}\} \\ 
-(\mathcal{E}-1)[h+i\frac{\Omega }{c}(\mathcal{E-E}_{0})\tilde{x}-\frac{%
\Omega }{c}(\mathcal{E-E}_{0})\tilde{y}]%
\end{array}%
\right) ,  \label{sd12} \\
N_{-12} &=&\exp [-\frac{d_{2}^{2}}{2d}]\frac{|\mathcal{E}_{0}|mc/\sqrt{2\pi }%
}{2\hbar \sqrt{\mathcal{E}^{2}h^{2}+(\mathcal{E-E}_{0})^{2}(1+4\hbar
^{2}d/m^{2}c^{2})}},  \label{Nd12}
\end{eqnarray}%
their eigenvalues satisfy the characteristic equation 
\begin{equation}
\mathcal{E}(\mathcal{E+}\Lambda _{\pm })-1-\varsigma -\frac{\mathcal{E}h^{2}%
}{(\mathcal{E}\mp \mathcal{E}_{0})}=0,\text{ \ }\varsigma =\frac{4\hbar ^{2}d%
}{m^{2}c^{2}}.  \label{E1}
\end{equation}%
"Plus solutions" $\psi _{+11},$ $\psi _{+12}$ may be obtained from $\psi
_{-11},$ $\psi _{-12}$ with help of the transformation $\psi _{+}\rightarrow
\alpha _{1}\alpha _{3}\psi _{-}$ and the simultaneous sign change of $%
\mathcal{E}_{0},\psi _{y}$ and $\mathcal{E}_{0},\psi _{x}$ for\ $\psi _{+11}$
and $\psi _{+12}$ respectively$.$

\section{3. Singular solutions}

Study an time-evolution of a state is described by the wave function
consisting from a sum of different wave functions. Except for the operator (%
\ref{inxy}), operators of energy, angular momentum, spin, etcetera don't
commute with Hamiltonian, therefore, average values of these operators have
to be used. The average value of any operator $P$ is%
\begin{equation}
\bar{P}=\sum_{k,l}\int C_{k}^{\ast }C_{l}\Psi _{k}^{\ast }P\Psi _{l}dxdy
\label{Pkl}
\end{equation}%
An oscillating term corresponds every pair of different wave functions. The
difference of any two roots of the characteristic equations defines the
oscillation frequency. Usually this frequency is large and cannot be
measured directly.

However, there exists a singular case where this frequency is always small.
In this case every root tends to $\pm \mathcal{E}_{0}$ at $h\rightarrow 0$.
Obviously the characteristic equations (\ref{E0}), (\ref{E1}), as equations
of third order, have exact solutions. However, since $H\ll |H_{z}|$ and $|%
\mathcal{E}_{0}|\sim 1,$ the normalized parameter $h\equiv \mathcal{E}%
_{0}H/H_{z}\ll 1$. Therefore, expansions of roots in terms of this parameter
is more convenient. The characteristic equations of the singular states have
a pair of roots in the form of the expansion in a vicinity of $\pm \mathcal{E%
}_{0}.$ Every such a root is expanded in power series in $h$. The third root
equals $-1/\mathcal{E}_{0}$ at $h=0$ and is expanded\ in power series in $%
h^{2}.$

The necessary condition for existence of singular states is%
\begin{equation}
\text{\ }\Lambda _{\mp }=\pm \frac{1}{\mathcal{E}_{0}}\mp \mathcal{E}_{0}.
\label{conl}
\end{equation}%
For simplicity we consider below "minus solutions" for $eH_{z}<0.$ For "plus
solutions" $eH_{z}>0$ the parameter $\mathcal{E}_{0}$ changes the sign.

Three roots of the ground state are 
\begin{equation}
\mathcal{E}_{1,2}^{\prime }=\mathcal{E}_{0}\pm \frac{\mathcal{E}_{0}h}{\sqrt{%
\mathcal{E}_{0}^{2}+1}}+\frac{\mathcal{E}_{0}h^{2}}{2(\mathcal{E}%
_{0}^{2}+1)^{2}}+\ldots ,\text{ \ }\mathcal{E}_{3}=-\frac{1}{\mathcal{E}_{0}}%
-\frac{\mathcal{E}_{0}h^{2}}{(\mathcal{E}_{0}^{2}+1)^{2}}+\ldots ,
\label{rg}
\end{equation}%
For the non-degenerate solutions of the first excited state the form of
roots coincides with (\ref{rg}), but in the condition (\ref{conl}) the term $%
\Lambda _{\mp }$ must be replaced by the expression (\ref{la}).

For degenerate solutions of the first "excited state" roots are 
\begin{equation}
\mathcal{E}_{1,2}^{\prime \prime }=\mathcal{E}_{0}\pm \frac{\mathcal{E}_{0}h%
}{\sqrt{\mathcal{E}_{0}^{2}+1+\varsigma }}+\frac{\mathcal{E}_{0}(1+\varsigma
)h^{2}}{2(\mathcal{E}_{0}^{2}+1+\varsigma )^{2}}+\ldots ,\text{ \ }\mathcal{E%
}_{3}=-\frac{1+\varsigma }{\mathcal{E}_{0}}-\frac{\mathcal{E}%
_{0}(1+\varsigma )h^{2}}{(\mathcal{E}_{0}^{2}+1+\varsigma )^{2}}+\ldots ,
\label{rdg}
\end{equation}%
and the same condition (\ref{conl}) is used.

\section{4. Phase matching condition}

The condition $\Lambda _{\mp }=0$ for singular states is the phase matching
between momentum of the wave and spinor 
\begin{equation}
\hbar \Omega =2\varepsilon pc.  \label{pm}
\end{equation}%
In accordance with Eq. (\ref{conl}), this condition corresponds to equality $%
\mathcal{E}_{0}=\pm 1.$ Using expressions for $d,\mathcal{E}_{0}$ in
non-normalized parameters (\ref{vh}), (\ref{d2}), it is easily shown that
the equality is none other than the classical condition of the magnetic
resonance 
\begin{equation}
\hbar \Omega =\pm \frac{e\hbar }{2mc}H_{z},  \label{cmr}
\end{equation}%
for the "normal magnetic moment" of electron.

In general case, the normalized parameter $\mathcal{E}_{0}$ defines the $g$%
-factor $\mathcal{E}_{0}=2/g.$ Quantum field theory produces a coefficient
at the first term in the right part of Eq. (\ref{conl}). As a result, $%
\mathcal{E}_{0}$ is changed and the magnetic moment becomes anomalous one.
But for simplicity, as it is noticed in Introduction, the consideration here
is restricted by "pure quantum mechanics".

Every term in (\ref{Pkl}) has multiplier $\exp \left[ -(d_{2}^{\prime
}-d_{2}^{\prime \prime })^{2}/2d\right] ,$ where $d_{2}^{\prime
},d_{2}^{\prime \prime }$ correspond to two eigenvalues of $\mathcal{E}.$ In
particular, for singular pair of the ground state this factor in the first
approximation equals $\exp \left[ -(\mathcal{E}_{0}^{2}+1)/2\lambda ^{2}d%
\right] ,$ where $\lambda $ is the Compton wavelength. For electron $\lambda
^{2}d=1.13\cdot 10^{-14}H_{z}/G,$\ therefore, this factor equals zero with
huge accuracy for all magnetic fields attained in laboratory conditions. The
same is valid for singular pair of excited states. In contrast to that, for
a sum of the ground and first excited degenerative state this factor is $%
\exp [-2d\lambda ^{2}].$ It equals one with huge accuracy. Therefore, for an
illustration of the temporal behavior of the average value, it is suffice to
consider this combination of wave functions.

Consider the average value of spin $\bar{s}=-i\frac{1}{2}\int \Psi ^{\ast
}\alpha _{1}\alpha _{2}\Psi dxdy$ for such a mixed state with the wave
function $\Psi =(C_{0}\Psi _{-0}+C_{1}\Psi _{-11})$. Without loss
generality, constants $C_{0},C_{1}$ are assumed to be real: $C_{0}=\sin \tau
,C_{1}=\cos \tau .$ It may be straightforwardly shown that the constant part
of $\bar{s}$ \ is negligible, because it is a sum of two small terms of the
order of $h$ or $\varsigma $. The oscillation term at $\mathcal{E}_{0}^{2}=1$
is 
\begin{equation}
\bar{s}(t)=\frac{1}{\sqrt{8}}\sin 2\tau \cos \digamma t,\text{ \ }\digamma =(%
\mathcal{E}^{\prime }-\mathcal{E}^{\prime \prime })mc^{2}/\hbar \approx 
\frac{eH}{mc\sqrt{8}},  \label{F}
\end{equation}%
where $\digamma $ is the oscillation frequency$.$

Basic parameters applicable to measurements are determined as follows. The
allowable value of $H_{z}$ is defined from Eq. (\ref{d2}) by the appropriate
transverse localization length $\sim 1/\sqrt{2d}$; some examples are
presented in \cite{S1}. $\Omega $ is defined by the condition (\ref{cmr}),
in particular, it may be a optical frequency \cite{S1}. The value of the
"momentum in the rotating and co-moving frame of references" $|$ $p|=|\hbar
\Omega /2c|$ is small in contrast to the average momentum $\bar{p}\equiv
\int \Psi ^{\ast }(-i\hbar \frac{\partial }{\partial z})\Psi dxdy.$ This
momentum with the accuracy of the order of $h$ equals $\varepsilon mc$; in
this expression $p$ is neglected. An average velocity $\bar{v}$ associated
with the average momentum $\bar{p}$ is $\bar{v}=c/\sqrt{2}$. This value
falls in the diapason of relativistic velocities.\ \ \ \ \ \ \ \ \ \ \ \ \ \
\ \ \ \ \ \ \ \ \ \ \ \ \ \ \ 

\section{5. Conclusion}

Singular states pertaining to the new class of exact solutions of Dirac's
equation in the rotating electromagnetic field are most practicable. They
may be separated from totality of states with help of the phase matching
condition. The use both the magnetic resonance and the resonance velocity
gives a chance for an improvement of the precision measurements of the
fermion mass and magnetic moment in the relativistic range.

\end{document}